\documentclass[osajnl,twocolumn,showpacs,superscriptaddress,10pt]{revtex4-1} 
\usepackage{amsmath,amssymb,graphicx}

\usepackage{fancyhdr}
\usepackage{lipsum}
\usepackage{array}

\begin{document}

\title{Anamorphic transformation and its \\  application to time-bandwidth compression} 

\author{Mohammad H. Asghari} \email{Corresponding author: asghari@ucla.edu}
\affiliation{Department of Electrical Engineering, University of California Los Angeles CA 90095, USA}
\author{Bahram Jalali}
\affiliation{Department of Electrical Engineering, University of California Los Angeles CA 90095, USA}
\affiliation{Department of Bioengineering, University of California, Los Angeles, CA 90095, USA}
\affiliation{Department of Surgery, David Geffen School of Medicine, University of California, Los Angeles, CA 90095, USA}

\begin{abstract}A general method for compressing the modulation time-bandwidth product of analog signals is introduced and experimentally demonstrated. As one of its applications, this physics-based signal grooming performs feature-selective stretch, enabling a conventional digitizer to capture fast temporal features that were beyond its bandwidth. At the same time, the total digital data size is reduced. The compression is lossless and is achieved through a same-domain transformation of the signal’s complex field, performed in the analog domain prior to digitization. Our method is inspired by operation of Fovea centralis in the human eye and by anamorphic transformation in visual arts. The proposed transform can also be performed in the digital domain as a digital data compression algorithm to alleviate the storage and transmission bottlenecks associated with "big data".  \\ \\
Keywords: Anamorphic stretch transform; Photonic time stretch, Time-stretch dispersive Fourier transform; Time stretch analog-to-digital conversion; Feature selective sampling; feature-selective time stretch; Same-domain transform; Warped dispersive Fourier transform; Generalized Dispersive Fourier Transform; Ambiguity function, Wigner distribution function; Warped temporal imaging; Generalized time-wavelength mapping; Time frequency distribution, Ultrafast processing; digital data compression; big data. 
\end{abstract}


\maketitle 
\thispagestyle{fancy}


\section{Introduction}

In conventional sampling, the analog signal is sampled at twice the highest frequency of the signal, the so-called Nyquist rate. This makes inefficient use of the available samples because frequency components below the Nyquist rate are oversampled. This uniform, frequency-independent sampling causes two predicaments: (i) it limits the maximum frequency that can be captured with a given sampling rate (to half of the sampling rate) and (ii) when the signal has redundancy, it results in a record length that is much larger than necessary (since low frequencies are oversampled). 

\begin{figure}[htbp]
\centerline{\resizebox{0.48\textwidth}{!}{\includegraphics[width=.8\columnwidth]{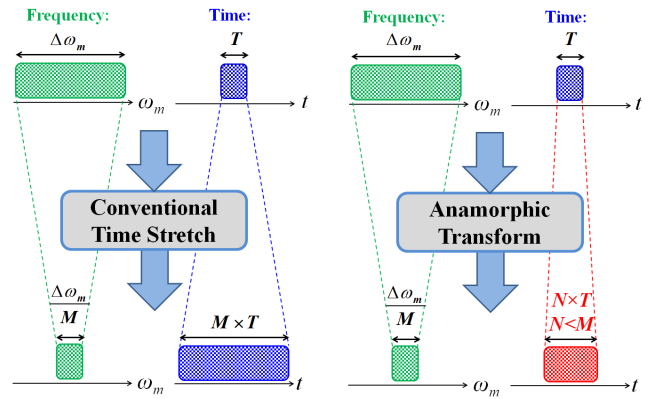}}}
\caption{Comparison of the conventional time-stretch transform (left) and proposed anamorphic transform (right). Both are performed prior to sampling and they boost the ADC’s sampling rate. However, for a given bandwidth compression factor $M$, the anamorphic transform leads to a shorter record length with fewer samples. ADC: Analog to Digital Converter. $\omega_m$ is the modulation frequency.}
\end{figure}

Time-stretching performed in the analog domain prior to sampling [1-7] overcomes the first problem by reducing the signal bandwidth (see Fig. 1). In this method, the signal is modulated on a chirped optical carrier and then subjected to Dispersive Fourier Transform (DFT), which causes the signal, now represented by the modulation intensity of the carrier, to be stretched in time (its bandwidth compressed). Since the photodiode measures the modulation intensity this reduces the bandwidth requirements of the photodiode and the analog to digital converter. Here the time-bandwidth product (TBP) remains constant because when the modulation intensity bandwidth is compressed by a factor M, the signal’s time duration is increased by the same factor. Fast features are suitably slowed down for the digitizer to sample and quantize them at the Nyquist rate; however, the slow features are oversampled. This redundant oversampling results in a needlessly large record length. It would be highly desirable to compress the bandwidth without this proportional increase in the time duration; in other words, a reduction of the modulation TBP. This requires feature-selective time-stretch. In principle, this should be possible when fast features occur sparsely. The benefit would be similar to that offered by compressive sensing [8,9], albeit, through warping of the signal as opposed to modification of the sampling process. 

In this paper, we propose and experimentally demonstrate a new transformation that compresses the time-bandwidth product of the modulation by reshaping the signal’s complex field in the analog domain before sampling and digitization. This variable sampling is performed by reshaping the signal with a phase filter with a nonlinear group delay. Because of the analogy to anamorphism in graphic arts (discussed later), we refer to this operation as the Anamorphic Transformation. We identify the specific group delay vs. frequency profile using the Modulation Intensity Distribution, a two-dimensional function that unveils the signal’s modulation bandwidth and its dependence on the group delay. The signal reshaping operation is then combined with complex field detection followed by digital reconstruction at the receiver. The net result is that the modulation bandwidth is reduced without the aforementioned expense of a proportional increase in temporal duration (see Fig. 1).  
 
Our technique makes it possible to capture an ultrafast signal in real-time with a digitizer that would otherwise have insufficient sampling rate. At the same time, the number of samples needed for digital representation, and hence the digital data size, is reduced. Our technique measures both the time domain and the spectrum of ultrafast signals in real-time. For application to optical waveforms, the nonlinear group delay filter operation can be performed with dispersive elements with engineered group velocity dispersion such as Chirped Fiber Bragg Gratings (CFBG), Chromo Modal Dispersion [10] or free space gratings.

While this paper focuses on applications in capturing ultrafast analog signals where the anamorphic transformation is performed in the analog domain, our mathematical transformation can also be performed in digital domain on digital data. This all-digital implementation would be a data compression algorithm that may prove useful in dealing with the storage and transmission bottlenecks of “big data”.

\subsection{Analogy with the Biological Eye}

By reshaping the signal prior to digitization, the proposed anamorphic transformation causes the digitizer’s uniformly spaced samples to be nonuniformly distributed.  Our method is inspired by the Fovea centralis in human eye. The Fovea centralis is a part of the eye located near the center of the retina. It has a much higher density of photoreceptors than the rest of the retina and is responsible for the high resolution of central vision, necessary for humans to read, watch, drive, and other activities where visual detail is of primary importance. While the Fovea comprises less than $1\%$ of the retina, it uses over $50\%$ of the brain’s visual processing power. Since photoreceptors perform sampling, the Fovea causes nonuniform sampling of the field of view. Although the physical sample density (sample rate of the digitizer) in our system is uniform, we achieve a non-uniform distribution of samples across the signal by reshaping the signal prior to sampling in the temporal domain. While there is no nonlinear filter in the eye, the eye achieves similar nonuniform sampling by using nonuniform photoreceptor density provide by the Fovea.  Hence our technique can be interpreted as biomimicry of the human eye. 

\subsection{Analogy with Anamorphism in Graphic Arts}

The reshaping of the signal in our technique evokes comparison to anamorphic image transformation techniques used to create optical illusion and art [11]. Fundamental differences exist between our technique and the conventional anamorphic imaging. First, our technique warps the frequency domain (Fourier domain), whereas in anamorphic imaging the image is warped in its original spatial domain. Second, the image transformation in anamorphic imaging is arbitrary and chosen for artistic considerations, or to change the aspect ratio of the image. In contrast, in our technique the transformation self-adapts to the frequency content of the signal causing fast time features to be slowed down more than the slower ones. This self-adaptation occurs naturally and is a consequence of the frequency dependence of optical dispersion (used to create the transformation as discussed later). Third, our technique works in the time domain, in particular it is applied to the digitization (e.g. ADC and DAC) and processing of temporal waveforms such as communication signals. Because of the non-uniform warping of signal’s spectrum in our technique, albeit executed in the frequency domain, our technique may be referred to as the Anamorphic Transform, or Warped Dispersive Fourier Transform.

\section{Technical Description}

A passband analog signal can be represented by an modulation (baseband) waveform modulated on a carrier. ADCs usually detect the modulation of the input signal, i.e. after downconversion. Here we derive a mathematical algorithm that describes the optimum analog transformation that reshapes the spectrum of the signal such that its modulation can be captured with an ADC that would otherwise be too slow. Unlike the conventional uniform time-stretch processing, the new transformation minimizes the record length and number of samples. This transformation is implemented via a filter with an engineered group delay. 

\begin{figure}[htbp]
\centerline{\resizebox{0.48\textwidth}{!}{\includegraphics[width=.8\columnwidth]{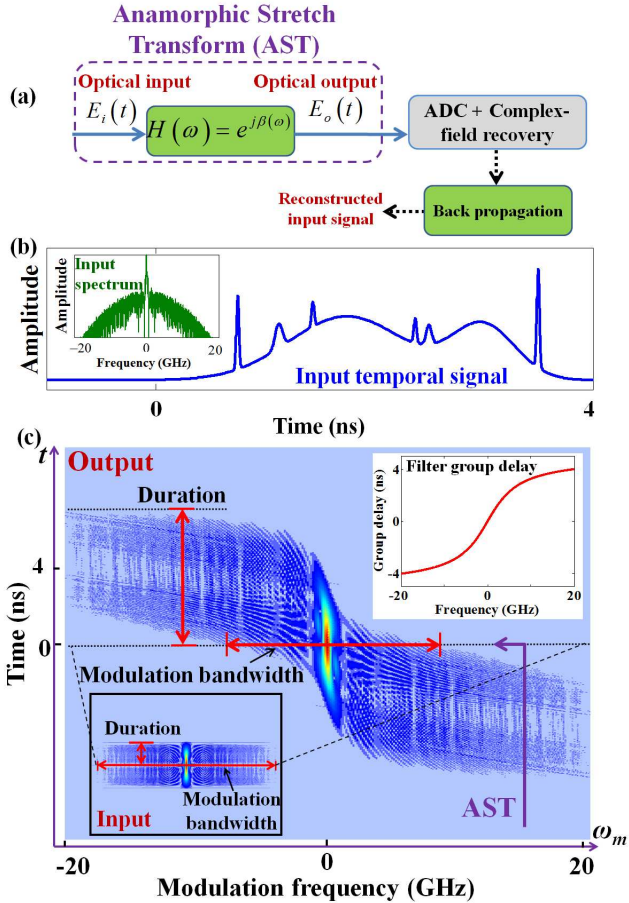}}}
\caption{(a) The proposed anamorphic transformation is performed using a filter with a tailored frequency-dependent group delay placed prior to the analog-to-digital converter (ADC). The complex field of the transformed signal is measured and the input signal is reconstructed using back propagation. (b) An arbitrary input signal;  inset shows its field spectrum. (c) The Modulation Intensity Distribution (MID) of the signal after it is subjected to a filter with S-shaped group delay (see inset of (C)). The MID is a 3D plot showing dependence of the modulation intensity (color) on time and modulation frequency. For comparison, the MID of the input signal without the filter is shown in the inset. The anamorphic transform reduces the signal modulation bandwidth but it does not lead to a proportional increase in its time duration. The complex interference patterns arise because the system is in the near field. $\omega_m=0$ corresponds to the carrier frequency.}
\end{figure}

Temporal group velocity dispersion can be represented by a filter with quadratic phase, i.e. one with the transfer function  $H(\omega)=e^{j.\beta_2.\omega^2/2}$. We generalize the problem by allowing the phase to be an arbitrary function of frequency (see Fig. 2(a)). $H(\omega)$ is the spectral response of a filter with phase $\beta(\omega)$ and group delay (GD) of $\tau(\omega)= \partial[\beta(\omega)]/\partial\omega$. The complete list of parameters and acronyms we have used in this paper is given in Table 1 in the Appendix. The modulation intensity spectrum of the input signal can be described in terms of the complex amplitude $E_i(t)$:

\begin{equation}
I_i(\omega_m)=FT\{|E_i(t)|^2\}
\end{equation}

where $FT\{\}$ is the Fourier transform operator and $\omega_m$ is the modulation (sideband) frequency measured with respect to the carrier frequency, $\omega$. It is easy to show that the modulation spectrum can be written as a correlation function:

\begin{equation}
I_i(\omega_m)=\int\limits_{-\infty}^\infty \tilde{E}_i(\omega) \tilde{E}_i^*(\omega+\omega_m) d\omega
\end{equation}

where  $\tilde{E}_i(\omega)$  is the spectrum of the input signal. Equation (2) describes the correlation of the electric field with its frequency-shifted copy and calculates the spectrum of the modulation intensity. After passing through the filter, the modulation spectrum of the output signal, can be calculated as follow:

\begin{equation}
I_o(\omega_m)=\int\limits_{-\infty}^\infty \tilde{E}_i(\omega) \tilde{E}_i^*(\omega+\omega_m) e^{j(\beta(\omega)-\beta(\omega+\omega_m))} d\omega
\end{equation}

Here we define a new transform called Anamorphic Stretch Transform (AST) [12], that relates the input \emph{carrier} (field) spectrum to the output \emph{modulation} spectrum ($FT{|E_o(t)|^2}$): 

\begin{equation}
\begin{split}
\text{AST}\{\tilde{E}_i(\omega)\}(\omega_m)=\int\limits_{-\infty}^\infty \tilde{E}_i(\omega) \tilde{E}_i^*(\omega+\omega_m) \qquad \qquad \\ e^{-j \omega_m [\frac{\beta(\omega+\omega_m)-\beta(\omega)}{\omega_m}]} d\omega
\end{split}
\end{equation}

As seen later, for time-bandwidth compression, the shape of the optimum group delay is a sublinear function resembling the letter "S". Therefore, one may refer to this particular implementation as the S-Transform (ST), although it should be noted that the Anamorphic Stretch Transform is more general than this particular group delay function. 

For filters operating in the far field (i.e. filters with large group velocity dispersion (GVD)), the signal is stretched in time by a large amount, hence its modulation frequency  and the bracketed term in the exponent of equation 4 is reduced to the group delay,  $d\beta(\omega)/d\omega=\tau(\omega)$. Thus, in the far field condition, AST is simplified to:

\begin{equation}
\text{AST}\{\tilde{E}_i(\omega)\}(\omega_m)=\int\limits_{-\infty}^\infty \tilde{E}_i(\omega) \tilde{E}_i^*(\omega+\omega_m) e^{-j \omega_m \tau(\omega)} d\omega
\end{equation}

\subsection{Modulation Intensity Distribution (MID)} 
Anamorphic transform gives the modulation spectrum of the signal at the output of the filter. Since our objective is to simultaneously minimize the modulation bandwidth and time duration, we require a mathematical tool that describes both the modulation spectrum and its temporal duration. The following 2D distribution describes the modulation intensity spectrum and its dependence on time. We refer to this as the Modulation Intensity Distribution (MID):

\begin{equation}
\begin{split}
\text{MID}(\omega_m,t)=\int\limits_{-\infty}^\infty \tilde{E}_i(\omega) \tilde{E}_i^*(\omega+\omega_m) \qquad \qquad  \qquad \\ e^{-j \omega_m [\frac{\beta(\omega+\omega_m)-\beta(\omega)}{\omega_m}]}  e^{j \omega t} d\omega
\end{split}
\end{equation}

The modulation spectrum and time duration of a signal subject to an arbitrary group delay is obtained from this 2D distribution. This information is then used to design a filter with the right group delay. The MID can be mathematically described as the cross-correlation of the output signal spectrum with its temporally shifted waveform. At $t = 0$ (i.e. time shift of zero) the MID becomes the autocorrelation of the output signal spectrum (i.e. the output modulation spectrum). Thus the trajectory at $t = 0$ in the MID represents the output modulation spectrum (i.e. AST) and its width determines the output modulation bandwidth. Also the maximum absolute amount of the temporal shift that the cross-correlation has non-zero values is given by the time duration of the output signal. Hence, the output signal duration can be measured from the MID as half of the time range over which the MID has non-zero values. 

The MID plot for an arbitrary signal (see Fig. 2(b)) subjected to a filter with an S-shaped GD (see inset of (C)) is shown in Fig. 2(c). MID belongs to a class of time-frequency distribution functions that also includes the "Ambiguity function" [15] and "Wigner distribution". The 3D plot shows the dependence of the modulation amplitude (color) at the output on time and modulation frequency. The inset shows the same for the input signal. It relates the bandwidth and temporal length of the modulation to the filter phase response (group delay profile). By choosing the proper filter, we can engineer the modulation bandwidth of the signal to match the sampling rate of the ADC and its time duration to minimize the number of samples needed to represent it. As an example, the horizontal arrow shows the modulation bandwidth and the vertical arrow designates the time duration. It should be mentioned that the output signal has both amplitude and phase information requiring complex-field detection. The time domain signal can then be reconstructed from the measured complex field. A number of complex field detection techniques can be employed here [16 20].

While filters with arbitrary GD profiles can be considered for AST operation, here we are particularly interested in filters with general GD profiles that compress the modulation TBP. As suggested by the MID plot in Fig. 2, such filters should have a sublinear group delay profile. The $tan^{-1}$ function provides a simple mathematical description of such filters:

\begin{equation}
\begin{split}
\tau(\omega)=A . tan^{-1}(B.\omega),
\end{split}
\end{equation}

where $A$ and $B$ are arbitrary real numbers. Using Eq. (7), a wide range of filter GD profiles can be generated requiring only two parameters to represent them (see Section 5 for more information). Parameter $A$ in Eq. (7) is the amount of group delay dispersion and determines whether the filter is in the near field or far field regime. In the near field, $A$ is on the order of the input signal duration, whereas in the far field, $A$ is much larger than the duration. Parameter $B$ is related to the degree of anamorphism. Section 5 provides more information about the choice of $tan^{-1}$ function. 

The MID function shows that the modulation bandwidth is given by a trajectory through $t = 0$ of the MID, that is the horizontal axis. This property deserves an explanation as it is central to the utility of this new distribution function in identifying the optimum filter (group delay profile) that compresses the time-bandwidth product. The filter applies a phase shift that is an increasing function of frequency. Referring to Eq. (3), higher frequencies in the argument of the integral become highly uncorrelated and the integral over these fast oscillations vanishes. Thus the modulation bandwidth is governed by the central portion of the MID. Mathematically, this property is similar to the stationary phase approximation in dispersive Fourier transform [21,22].

Note that the modulation bandwidth defined in the MID (Fig. 2) is the passband (double sideband) bandwidth whereas after the photo detector, we would be concerned with the baseband (single sideband) bandwidth which would be half of the former.

\subsection{Comparison with Time-stretch Dispersive Fourier Transform}
Time-stretch Dispersive Fourier Transform (DFT) [3,17,21] can be considered a special case of the anamorphic transformation. In other words, in the limit when the group delay is linear, the system operates in the far field and detector performs intensity detection, anamorphic transform and dispersive Fourier transform are related via Fourier transform. Anamorphic transform, AST, can be described as generalized dispersive Fourier transform.

DFT relies on linear GVD to perform time dilation and Fourier transformation on the input signal in real time (Fig. 3(a)) [3]. DFT relates the input \emph{carrier} (field) spectrum to the output \emph{modulation} in the time domain through the following transformation:

\begin{equation}
\text{DFT}\{\tilde{E}_i(\omega)\}(t)=\left|\int\limits_{-\infty}^\infty \tilde{E}_i(\omega) e^{j \frac{\beta_2\omega^2}{2}} e^{j \omega t} d\omega \right|^2
\end{equation}

There are important differences between AST and DFT. First, DFT maps the input field spectrum to the output modulation in the time domain but AST maps the input field spectrum to the output modulation in the spectral domain (compare Eqs. (4) and (8)). Second, DFT occurs in the far field only whereas AST spans both far field and near field (see Section 3). Third, in DFT the filter has quadratic phase profile but in AST the filter has arbitrary phase profile. In other words, AST can be described as warped dispersive Fourier transform that also spans the near field [12]. 

In the limit when the filter has a quadratic phase profile, $\beta(\omega)=\beta_2.\omega^2/2$ with large phase change, i.e. $\beta_2>>T^2/{8\pi}$, AST is related to DFT through Fourier transform. Here $T$ is the duration of input signal. This case refers to time stretch DFT, which is a special case of the Anamorphic Stretch Transform. 

\begin{figure}[htbp]
\centerline{\resizebox{0.48\textwidth}{!}{\includegraphics[width=.8\columnwidth]{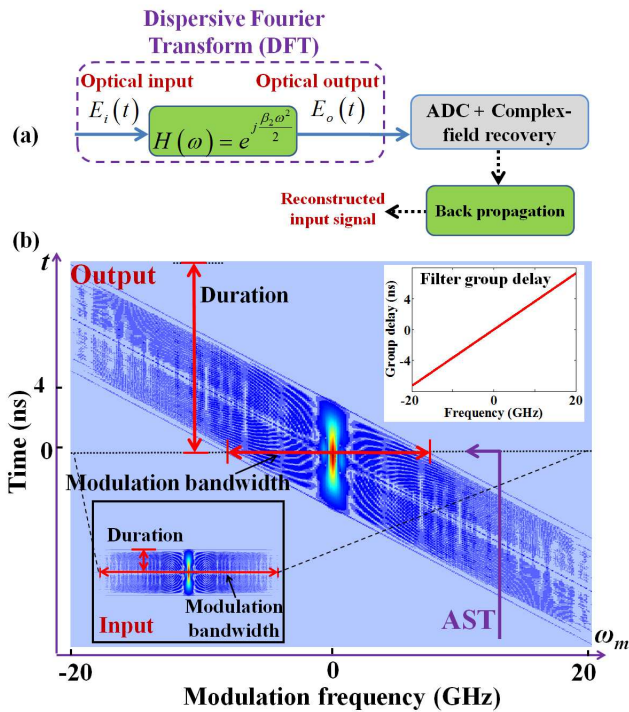}}}
\caption{(a) Anamorphic transformation in the case of a filter with a quadratic phase profile (linear group delay), i.e. the case of Dispersive Fourier Transform (DFT). (b)  Modulation Intensity Distribution (MID). The input signal is same as that in Fig. (2). $\omega_m=0$ corresponds to the carrier frequency.}
\end{figure}

When the MID is applied to a system with linear GD (quadratic phase), it provides valuable insight into how dispersion affects the time bandwidth product of signals in such a system.  Fig. 3(b) is the MID for such a system. To show the analytical power of the MID, we have considered a system in the near field.

Time stretch DFT has been shown to be a powerful method for real-time high-throughput spectroscopy [22-24] and imaging [25 27]. Owing to their high-throughput streaming operation, these instruments generate massive amounts of data, the storage and processing of which becomes challenging. Compared to DFT, the proposed anamorphic transformation reduces the record length and hence digital data size, easing the problem of big data in DFT-based real-time instruments. 

\subsection{Comparison with Photonic Time Stretch}
Photonic time stretch [1-7] is a DFT based method that compress the input signal (modulation) bandwidth so it can be captured using a photodiode and analog-to-digital converter that otherwise would have insufficient bandwidth. With digital reconstruction, our method can capture an ultrafast signal, in the same spirit as the photonic time stretch system.

There are important differences between our method and conventional time stretch concept. First, in conventional time stretch the signal is modulated on a chirped optical carrier using a modulator (mixer) and then is subjected to large amount of dispersion causing the signal to be stretched in time (its bandwidth compressed). The present method does not use any modulator. Second, in conventional time stretch the modulation TBP does not change. This means that when the signal (modulation) bandwidth is compressed M times, its duration is also increased M times. In our method the modulation TBP is compressed. When the signal modulation bandwidth is compressed M times, its duration is not increased proportionally (as depicted in Fig. 1). In temporal imaging (see e.g. [28]), TBP is similarly conserved. The present technique can be used to realize a warped temporal imaging system with added benefit of TBP compression.

\section{Far Field Regime}

In the first example on engineering the MID, we discuss the optimum group delay (GD) profile for a filter operating in the far field condition. 

The far field and near field regimes of group velocity dispersion can be understood in terms of the stationary-phase approximation. The far field corresponds to having sufficient dispersion to satisfy the stationary phase approximation while the near field refers to the regime before the approximation is satisfied [3,21].

We aim to compress the modulation bandwidth of the input analog signal while minimizing its duration. As an example, we consider an input signal with modulation bandwidth of 1 THz (field bandwidth 0.5 THz) and duration of 180 ps, see Fig. 4(a). The MID of the input signal without any filter in the system is shown in Fig. 4(b). We aim to compress the input signal modulation bandwidth to 8 GHz, i.e. a compression factor of 125. 

\begin{figure}[htbp]
\centerline{\resizebox{0.48\textwidth}{!}{\includegraphics[width=.8\columnwidth]{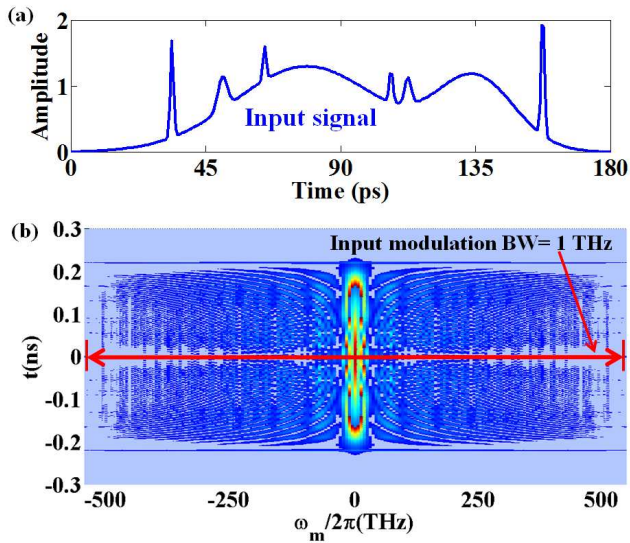}}}
\caption{(a) Input signal. (b) Modulation Intensity Distribution (MID) of the input signal without any filter. $\omega_m=0$ corresponds to the carrier frequency. Anamorphic transform of this input signal is shown in Fig. 6.}
\end{figure}

The filter transfer function is chosen such that GD for higher frequencies is less than the case of linear GD. This is because to achieve the same output modulation bandwidth, the GD required to compress the bandwidth of the high frequency portion of the spectrum is less, achieved using Eq. (7) with $A=7.86\times 10^{-9} s$  and $B=6\times 10^{-13} s$. Fig. 4 compares the nonlinear GD with a linear GD that would have resulted in the same 8 GHz output modulation bandwidth. Notice that the frequency axis in this figure shows the frequency deviation, i.e. filter's zero dispersion (origin in the plot) is at the input signal’s carrier frequency. 

\begin{figure}[htbp]
\centerline{\resizebox{0.48\textwidth}{!}{\includegraphics[width=.8\columnwidth]{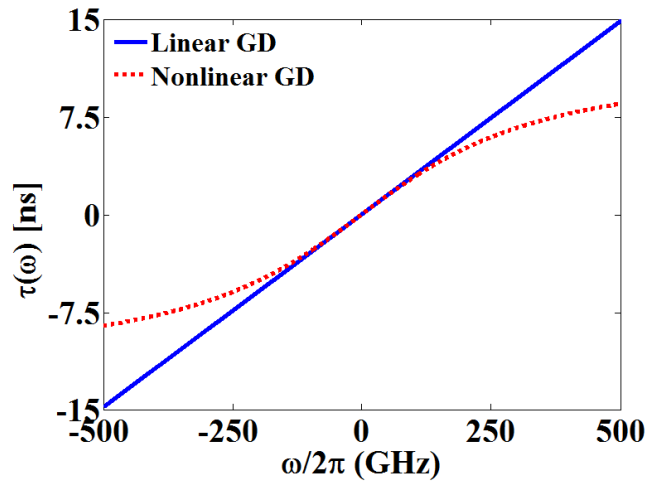}}}
\caption{Comparison of the linear and nonlinear filter Group Delay (GD) profiles that result in the same output modulation bandwidth. As observed in Fig. 6, the nonlinear GD results in a smaller time duration. $\omega=0$ corresponds to the carrier frequency.}
\end{figure}

\begin{figure}[htbp]
\centerline{\resizebox{0.48\textwidth}{!}{\includegraphics[width=.8\columnwidth]{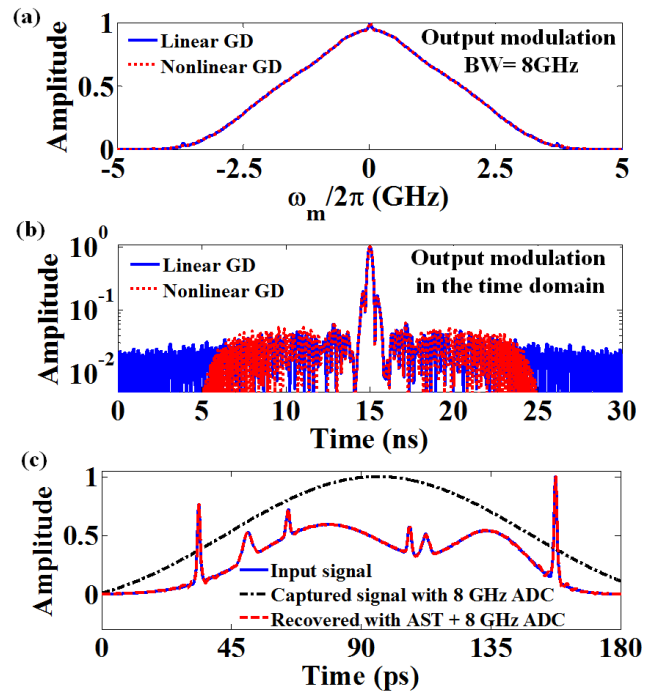}}}
\caption{Time-bandwidth compression using the Anamorphic Stretch Transform (AST) in the far field regime. (a) Comparison of the output modulation spectrums for filters with linear group delay (GD) (solid blue line) and with the tailored nonlinear GD, i.e. AST (dotted red line). Input signal is shown in Fig. 4 and filter GD profiles are shown in Fig. 5. (b) Comparison of the temporal outputs for the two filters. (c) Comparison of the recovered with the original signal. In both cases the modulation bandwidth is reduced from 1 THz to 8 GHz, however the temporal length, and hence the number of samples needed to represent it, is nearly $40\%$ lower with the AST. The signal captured with the same 8 GHz analog-to-digital converter (ADC) but without AST is also shown in (c). Modulation intensity distribution (MID) plots are shown in Fig. 7. $\omega_m=0$ corresponds to the carrier frequency.}
\end{figure}

As seen in Fig. 6(a), the modulation bandwidth is 8 GHz in both cases. However, the temporal duration (see Fig. 6(b)) is ~18 ns vs. 30 ns, i.e. $40\%$ reduction. Fig. 6(c) compares the recovered input signal using AST method (red dotted curve) with the input signal (blue solid curve). Captured signal with the same 8 GHz ADC but without AST is also shown with a black dash-dot curve. Fig. 6 shows that using AST the input signal can be captured accurately with an ADC that has lower bandwidth than the input signal. AST also minimizes the record length in comparison to the case of using a filter with linear GD.

Figure 7 compares the MID plots for the case of linear GD and the nonlinear GD used here. These MID plots were used to design and analyze the optimized bandwidth compression system in this example. The distribution is characterized by a well-defined, sharp, trajectory because the system is operating in the far field.

\begin{figure}[htbp]
\centerline{\resizebox{0.48\textwidth}{!}{\includegraphics[width=.8\columnwidth]{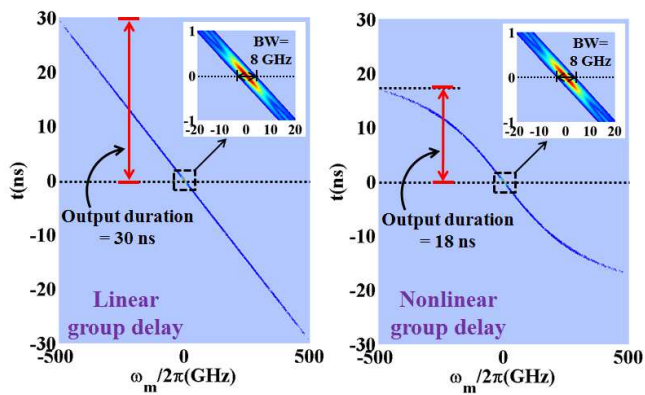}}}
\caption{Left and right figures show the Modulation Intensity Distribution (MID) of the signal in Fig. 6, when the filter has a linear and nonlinear (S-shaped) GD, respectively. In both cases the modulation bandwidth is reduced from 1 THz to 8 GHz, however the temporal length, and hence the number of samples needed to represent the signal, is nearly $40\%$ lower with the anamorphic transform. MID is used to identify the optimum GD profile. $\omega_m=0$ corresponds to the carrier frequency.}
\end{figure}

\section{Near Field Regime}

As another example we discuss the optimum GD profile for time-bandwidth compression using a filter operating in the near field. This would be important for cases where far field regime cannot be achieved because of insufficient available GD or limited bandwidth of the input signal. In this example, the input signal has an modulation bandwidth of 40 GHz (field bandwidth 20 GHz) and a 4 ns time duration (cf. Fig. 8(a)). The MID of the input signal is shown in Fig. 8(b). We aim to compress the input signal modulation bandwidth to 16 GHz, i.e. a compression factor of 2.5. 

\begin{figure}[htbp]
\centerline{\resizebox{0.48\textwidth}{!}{\includegraphics[width=.8\columnwidth]{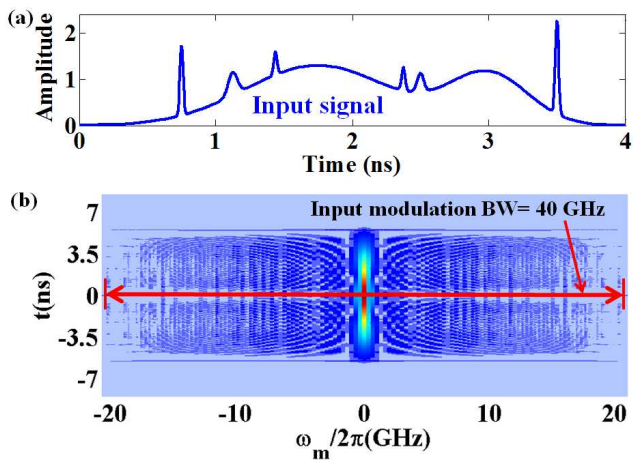}}}
\caption{(a) Input signal, (b) Modulation Intensity Distribution (MID) of the input signal. Anamorphic transform of this input signal is shown in Fig. 10. $\omega_m=0$ corresponds to the carrier frequency.}
\end{figure}

The filter transfer function is chosen such that for frequency components ranging from DC to 8 GHz a larger GD is applied to higher frequencies than the case of linear GD. The GD for frequency components above 8 GHz is designed to be less than the case of linear GD. This is because to achieve the same output modulation bandwidth, less group delay is required for fast features. Specifically, the chosen parameters for the filter’s group delay profile given by Eq. (7) is $A=3.14\times10^{-9} s$ and $B=2.7\times10^{-11} s$. Fig. 9 compares the nonlinear GD used with a linear GD that would have resulted in the same 16 GHz output modulation bandwidth.

\begin{figure}[htbp]
\centerline{\resizebox{0.48\textwidth}{!}{\includegraphics[width=.8\columnwidth]{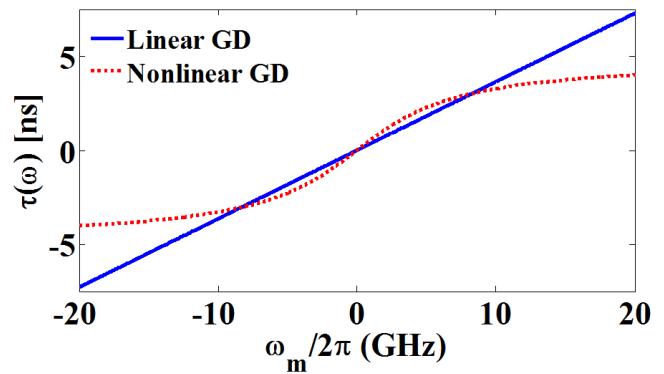}}}
\caption{Comparison of the linear and nonlinear filter Group Delay (GD) profiles that result in the same output modulation bandwidth. As observed in Fig. 10, the nonlinear GD results in a smaller time duration. $\omega=0$  corresponds to the carrier frequency.}
\end{figure}

\begin{figure}[htbp]
\centerline{\resizebox{0.48\textwidth}{!}{\includegraphics[width=.8\columnwidth]{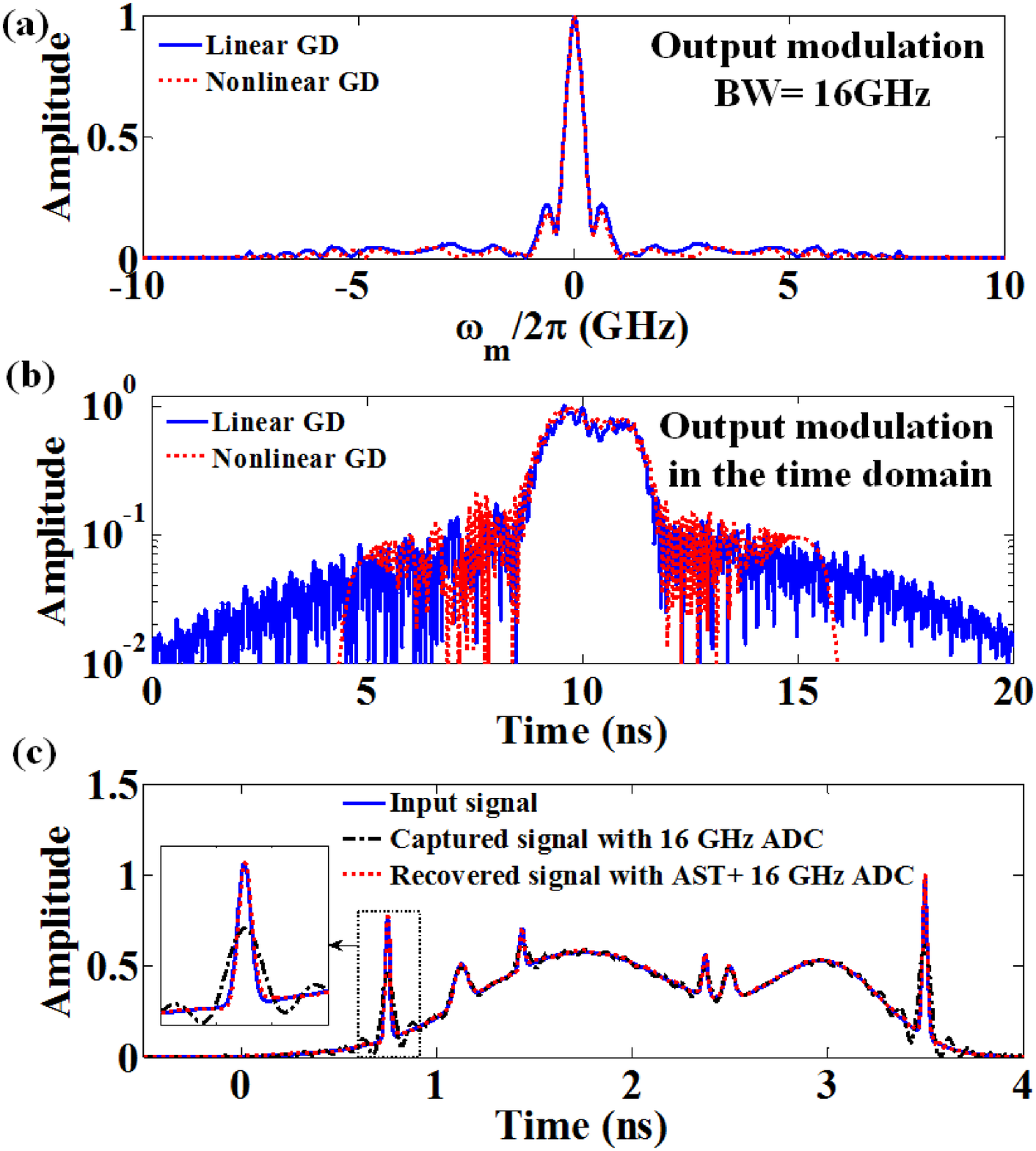}}}
\caption{Time-bandwidth compression using the Anamorphic Stretch Transform (AST) in the near field regime. (a) Comparison of the output modulation spectrums for filters with linear group delay (GD) (solid blue line) and with the nonlinear (S-shaped) GD (dotted red line). Input signal is shown in Fig. 8 and filter GD profiles are shown in Fig. 9. (b) Comparison of the temporal outputs for the two filters. (c) Comparison of the recovered signal with the original signal. In both cases the modulation bandwidth is reduced from 40 GHz to 16 GHz, however the temporal length, and hence the number of samples needed to represent the signal, is nearly $35\%$ lower with the AST. Captured signal with the same 16 GHz analog-to-digital converter (ADC) but without AST is also shown in (c). Modulation intensity distribution (MID) plots are shown in Fig. 11. $\omega_m=0$ corresponds to the carrier frequency.}
\end{figure}

\begin{figure}[htbp]
\centerline{\resizebox{0.48\textwidth}{!}{\includegraphics[width=.8\columnwidth]{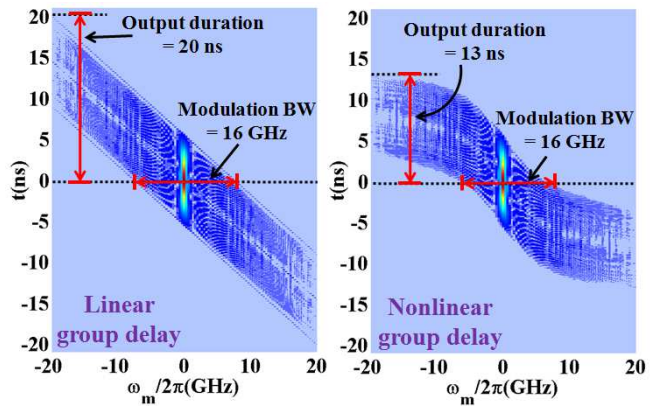}}}
\caption{Left and right figures show the Modulation Intensity Distribution (MID) of the signal in Fig. 10, when the filter has a linear or nonlinear (S-shaped) GD, respectively. In both cases the modulation bandwidth is reduced from 40 GHz to 16 GHz, however the temporal length, and hence the number of samples needed to represent it, is nearly $35\%$ lower with the anamorphic transform. MID is used to design the optimum GD profile.}
\end{figure}

As seen in Fig. 10(a), the modulation bandwidth is 16 GHz in both cases. However, the temporal duration (see Fig. 10(b)) is ~13 ns vs. 20 ns, i.e. $35\%$ reduction. Fig. 10(c) compares the recovered input signal using the AST method (red dotted curve) with the original input signal (blue solid curve), while the captured signal with the same 16 GHz ADC but without AST is also shown with black dash-dot curve. Fig. 10 shows that using AST input signal can be captured accurately with an ADC that has lower bandwidth than the input signal. AST also minimizes the record length for bandwidth compression in comparison to the case of using a filter with linear GD.  

Figure 11 compares the MID plots for the case of linear GD and the nonlinear GD used here. These plots were used to identify the optimum GD profile. The complex interference patterns in the MID plots arise because the system is operating in the near field.

\section{Discussion}

AST can be considered the generalized (or nonlinear) time-wavelength mapping. It reduces the modulation bandwidth so the signal can be captured with an ADC with a bandwidth that would otherwise be insufficient. At the same time, it minimizes the number of samples needed for a digital representation of the signal; in other words, it reduces the record length or the digital data size. A valid question is whether this time-bandwidth compression results in a loss of information. As a consequence of AST, a portion of the information contained in the signal modulation is transferred into the phase of the carrier. Hence no information is lost and the compression is lossless. Because some of the information is now contained in the phase, complex field detection is necessary in order to recover the original signal.

AST uses an all-pass filter to add phase shift to the input signal the amount of which increases with frequency in a prescribed manner. The proposed Modulation Intensity Distribution (MID) shows that, in order to compress the time-bandwidth product, the filter must have a nonlinear group delay profile, with proper slope at the origin (at carrier central frequency). The slope at the origin is inversely proportional to the modulation bandwidth. The relation between the filters with linear and nonlinear GDs can be represented by an all-pass filter with a rational polynomial function. In the region of interest, close to the origin, the lowest order polynomial gives the tan-1 function in Eq. (7). The proof of this is beyond the scope of this paper.

A tailored dispersion profile can be obtained by a number of techniques such as Chirped Fiber Bragg Grating (CFBG) with custom chirp [29], Chromo Modal Dispersion (CMD) [10] or diffraction gratings [30]. CFBG offers great flexibility in dispersion profile and low insertion loss. At the same time, it exhibits group delay ripples which are problematic. The recently demonstrated technique for mitigating these GD ripples [31] may be employed in our technique.  

\section{Experimental Demonstration}

We aim to compress the modulation bandwidth of the input analog signal while minimizing its duration. Fig. 12(a) shows one possible implementation of the AST system used in our experiments. The experimental setup used for demonstration of time-bandwidth compression is shown in Figs. 12(b) and (c). The experiments compare the time-bandwidth of the signal for (b) nonlinear (inverse tangent) group delay (GD) and (c) for linear group delay, and the results validate the time-bandwidth compression using the specifically designed nonlinear GD. The specific GD profile was obtained using the MID function, see Eq. (6). The nonlinear GD is experimentally realized using a custom chirped fiber Bragg grating, and the linear GD is realized using dispersion compensating fiber (DCF). To reconstruct the input signal from the measured waveform, output complex-field recovery is required followed by digital back-propagation technique. In this demonstration we used the STARS [16] technique for complex field measurement, although other complex field recovery methods may also be used. 

\begin{figure}[htbp]
\centerline{\resizebox{0.48\textwidth}{!}{\includegraphics[width=.8\columnwidth]{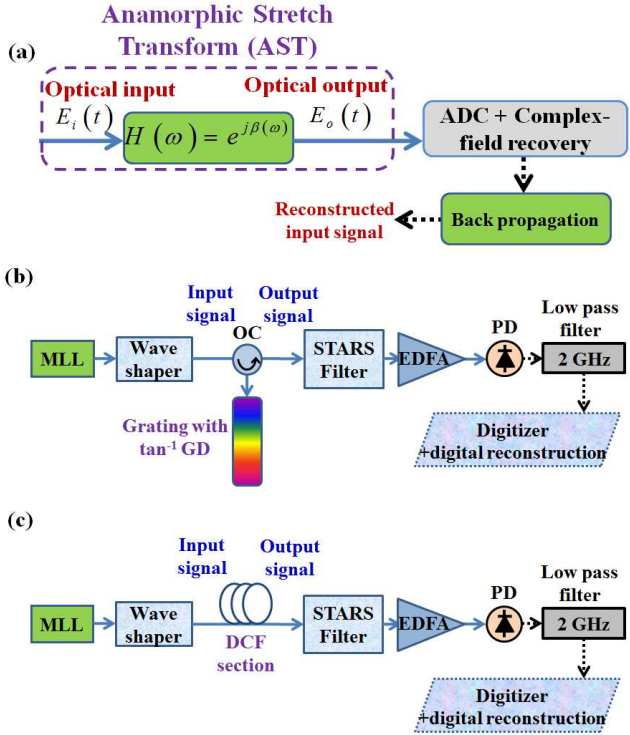}}}
\caption{(a) The Anamorphic Stretch Transform (AST) system. AST is a physics-based signal transformation that enables a digitizer to capture signals that would otherwise be beyond its bandwidth and at the same time, it compresses the digital data volume. This method is inspired by operation of Fovea centralis in the human eye and by anamorphic transformation in visual arts. AST makes it possible to (i) capture high-throughput random signals in real-time and (ii) to alleviate the storage and transmission bottlenecks associated with the resulting big data. It does so by compressing the time-bandwidth product. Experimental setup used for demonstration of time-bandwidth compression is shown in (b) and (c). The experiments compare the time-bandwidth of the signal for (b) a linear and nonlinear (inverse tangent) group delay (GD) and (b) for linear group delay. The nonlinear GD is realized using chirped fiber Bragg grating with nonlinear chirp. The linear GD is realized using dispersion compensating fiber (DCF). To reconstruct the input signal from the measured waveform, output complex-field recovery is required followed by digital back-propagation technique. In this demonstration we used the STARS [16] technique for complex field measurement. MLL: Mode-locked laser, OC: Optical circulator, EDFA: Erbium-doped fiber amplifier, PD: photodiode.}
\end{figure}

The test input signal was generated using Mode-Locked Laser (MLL) and an optical WaveShaper (Finisar 1000s). The input signal was designed using numerical simulations and had a field modulation bandwidth of 1 THz and duration of 50 ps, see Fig. 13. Its field spectrum is shown in the inset. We aim to compress the input signal electrical bandwidth from 1 THz to 2 GHz, i.e. a compression factor of 500. 

\begin{figure}[htbp]
\centerline{\resizebox{0.48\textwidth}{!}{\includegraphics[width=.8\columnwidth]{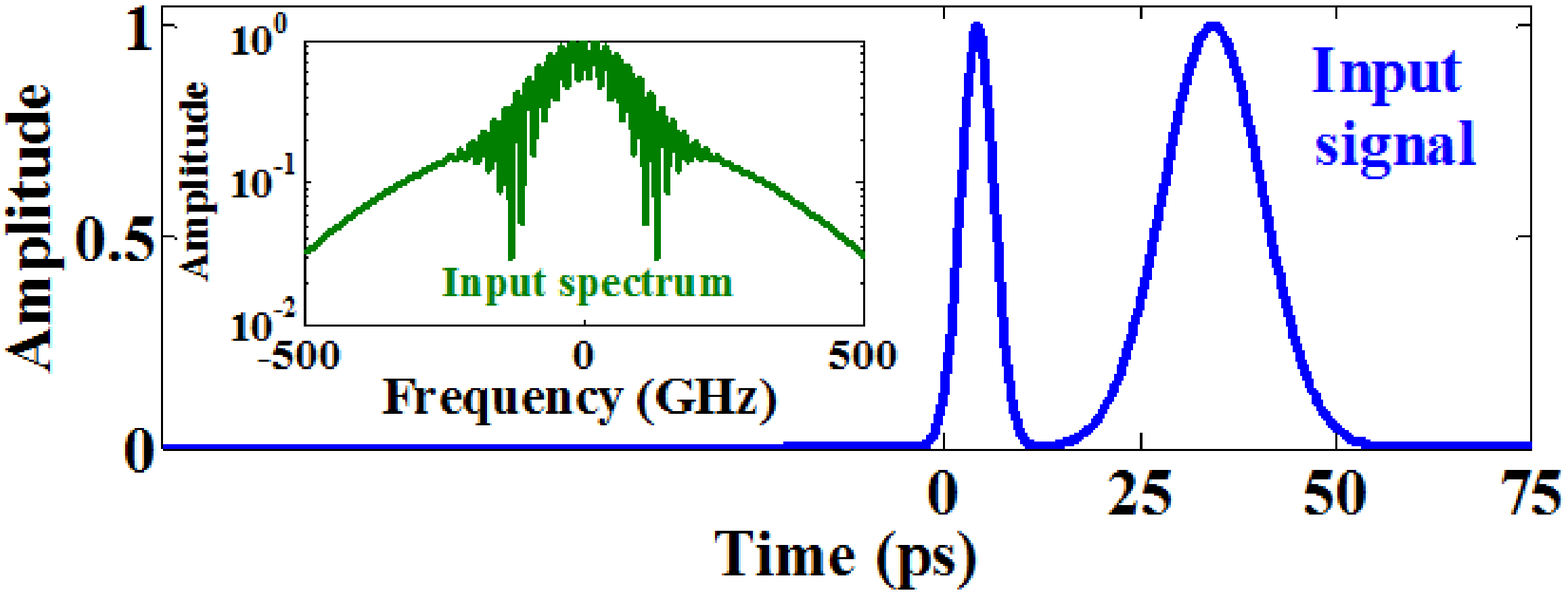}}}
\caption{The input signal. The signal was designed using numerical simulation. The complex spectrum obtained from simulation was programmed into the WaveShaper to produce the physical input to the experiment.}
\end{figure}

To show the effectiveness of our technique, we compare the case of AST using filters with linear GD to the case of nonlinear, specifically inverse tangent ($tan^{-1}$) GD. The linear case, in the far field, corresponds to the well-known time-stretch DFT. Fig. 14 shows comparison of the different filter GD used in the experiment. For time-bandwidth compression, the AST uses an inverse tangent profile. To show time-bandwidth compression, this results were compared to those that use linear GD (realized using dispersion compensating fiber (DCF) modules.  Two different modules were used: "Small GD" has total GD equal to that of AST filter, and "Large GD" has the same GD slope at the origin. Specifically, Large $GD = 25,600 ps^2$ and Small $GD = 8,800 ps^2$. For the inverse tangent group delay, we used the following profile: 

\begin{equation}
\begin{split}
\tau(\omega)=A . tan^{-1}(B.\omega),
\end{split}
\end{equation}

where $A=5 \times 10^{-9} s$ and $B=8.7 \times 10^{-13} s$. The AST filter with $tan^{-1}$ GD was implemented using a CFBG with customized grating chirp profile. Referring to Fig. 14, we note that the frequency axis is the frequency deviation, i.e. filter's zero dispersion (origin in the plot) is at the input signal’s carrier frequency. 

\begin{figure}[htbp]
\centerline{\resizebox{0.48\textwidth}{!}{\includegraphics[width=.8\columnwidth]{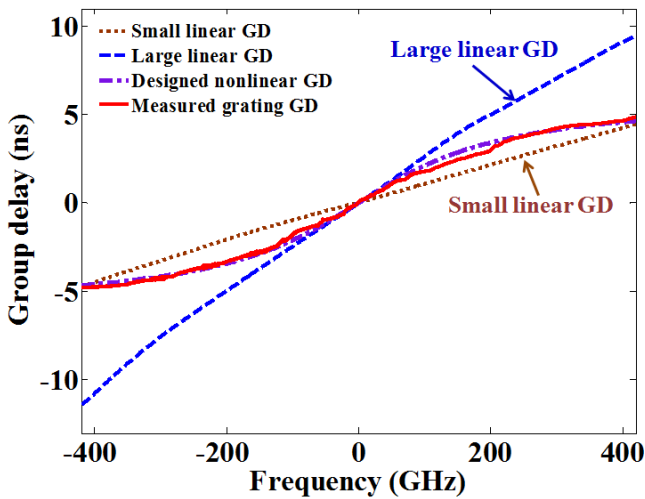}}}
\caption{Comparison of the different filter group delays (GD) used in the experiment. The Anamorphic Stretch Transform (AST) uses a sublinear GD, specifically an inverse tangent profile. This GD was realized in the experiments using a custom chirped fiber Bragg grating (CFBG). To show time-bandwidth compression, this results were compared to those that use linear GD (realized using dispersion compensating fiber (DCF) modules.  Two different modules were used: "Large GD" has total GD equal to that of AST filter, and "Small GD" has the same GD slope at the origin. Large GD = 25,600 $ps^2$; Small GD = 8,800 $ps^2$. $\omega = 0$ corresponds to the carrier frequency.}
\end{figure}

Fig. 15(a) compares the measured modulation intensity spectrums. In order to compare the true bandwidth of the waveforms, the 2 GHz low pass filter shown in Fig. 12 was not used in these measurements. As clearly seen, the electrical bandwidth in case of Small GD is 5.5 GHz and in cases of Large GD and $tan^{-1}$ GD is 2 GHz (i.e. the target electrical bandwidth). 

Fig. 15(b) compares the measured output temporal intensity profiles. Here, the 2 GHz electrical low pass filter emulates a system with only 2 GHz analog input bandwidth. In the case of Small GD, the output electrical bandwidth is 5.5 GHz so after the 2 GHz low pass filter, the measured signal has lost its higher frequency features. In cases of Large GD and $tan^{-1}$ GD, electrical bandwidths are reduced from 1 THz to the target 2 GHz.  However the temporal length, and hence the number of samples needed to represent the signal, is 2.73 times smaller with the $tan^{-1}$ GD. This clearly demonstrates time-bandwidth compression. 

Fig. 15(c) compares of the measured phase profiles. The dynamic range of the phase in the case of $tan^{-1}$ GD is less than that of for Large GD. 

It should be noted that this reduction in time duration results in higher peak power making the detection easier. Also, the in the case of Large GD, the loss of the dispersive element is about 18dB compared to about 1dB for the inverse tangent filter. In fact, to observe the signal in the case of Large GD, the signal had to be averaged 4000 times. Therefore, while the Large GD results is equally effective in reducing the electrical bandwidth, it has much lower signal to noise ratio than the inverse tangent GD. 

\begin{figure}[htbp]
\centerline{\resizebox{0.48\textwidth}{!}{\includegraphics[width=.8\columnwidth]{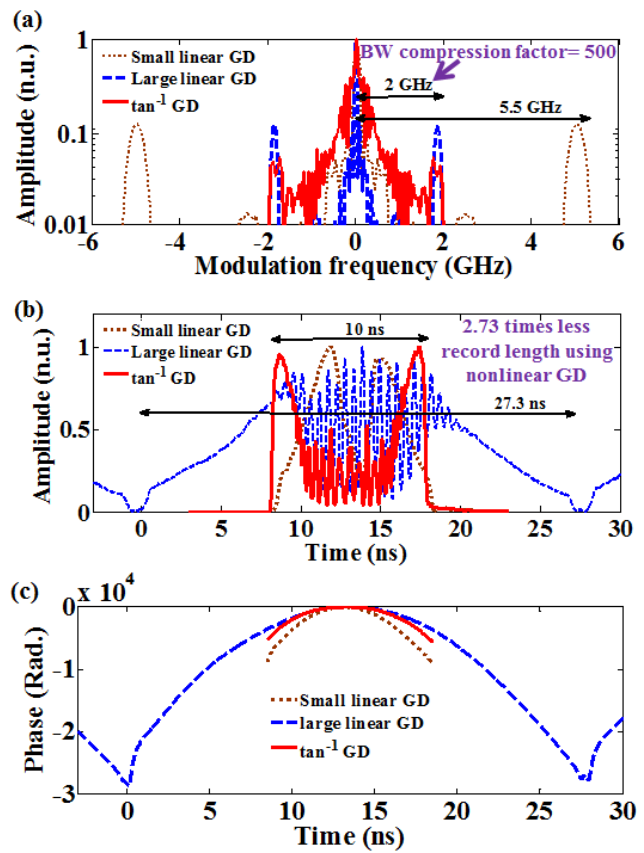}}}
\caption{Demonstration of time-bandwidth compression using the Anamorphic Stretch Transform (AST). (a) Comparison of the measured modulation spectrums with linear group delay (GD) (dotted brown and dashed blue lines) and with $tan^{-1}$ GD (solid red line). Input signal is shown in Fig. 13 and filter GD profiles are shown in Fig. 14. (b) Comparison of the measured temporal outputs after photo detection and the 2 GHz electrical low pass filter. In the case of Small GD, the modulation bandwidth is 5.5 GHz, so after the 2 GHz low pass filter, fast features are lost. In cases of Large GD and $tan^{-1}$ GD, bandwidths are reduced from 1 THz to the target 2 GHz, however the temporal length is 2.73 times smaller with $tan^{-1}$ GD. (c) Comparison of the measured phase profiles. The dynamic range of phase in the case of $tan^{-1}$ GD is less than that for Large GD.  n.u.: normalized unit.}
\end{figure}

For complex-field recovery, we have used STARS technique [16]. Spectral amplitude and phase response of the STARS filter is shown in Fig. 16(a). Fig 16(b) shows the oscilloscope traces of the two STARS outputs, for the $tan^{-1}$ GD. The complex-field is recovered using these two measurements and the algorithm we proposed in [16]. Using the complex field, the input signal is reconstructed via digital back propagation. Fig. 16(c) compares the recovered input signal with the original signal programmed into the WaveShaper.  In the case of Small GD, the input signal cannot be recovered because fast features are lost. In cases of Large GD and $tan^{-1}$ GD the input signal is properly recovered, however the temporal record length is 2.73 times lower with the $tan^{-1}$ GD. As noted earlier, the large losses in the case of Large GD necessitated signal averaging. Therefore, the results for this case are not real-time. 

\begin{figure}[htbp]
\centerline{\resizebox{0.48\textwidth}{!}{\includegraphics[width=.8\columnwidth]{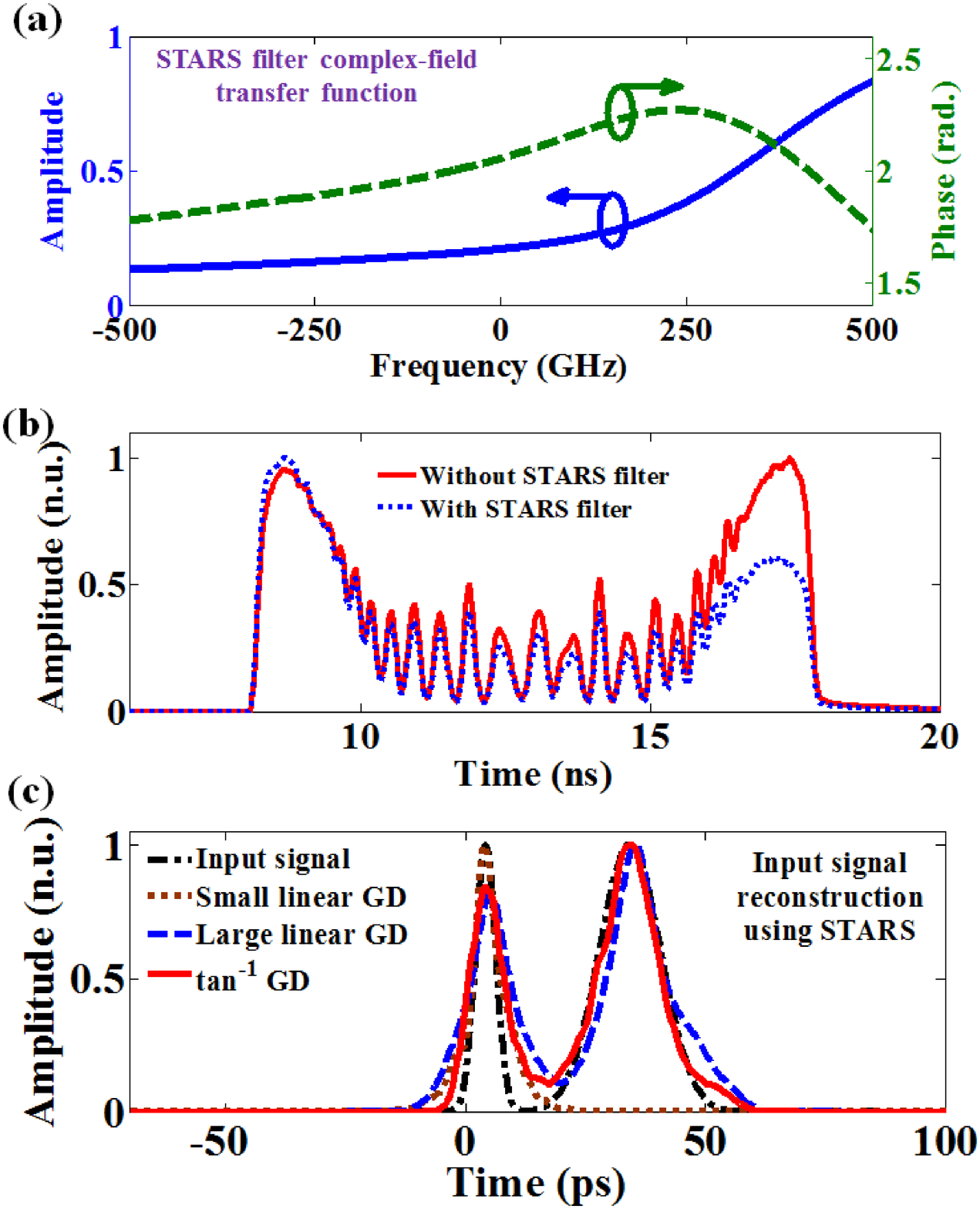}}}
\caption{(a) Spectral amplitude and phase response of the STARS filter used for complex-field recovery. $\omega = 0$ corresponds to the carrier frequency. (b) shows the oscilloscope traces of the two STARS outputs, for the $tan^{-1}$ GD. The complex-field is recovered using these two measurements. Using the complex field, the input signal is reconstructed via digital back propagation. (c) compares the recovered input signal with the original signal programmed into the WaveShaper.  In the case of Small GD, the input signal cannot be recovered because fast features are lost. In cases of Large GD and $tan^{-1}$ GD the input signal is properly recovered, however the temporal record length is 2.73 times lower with the $tan^{-1}$ GD. n.u.: normalized unit.}
\end{figure}

\section{Conclusions}
In this work we introduced a new mathematical transform that can be used to compress the modulation time-bandwidth of signals. This analog grooming is performed prior to digitization and is aimed (i) to overcome the bandwidth limitation of data converters (ii) to reduce the digital record length, and (iii) to enable real-time digital processing. Unlike in traditional time-stretching, the bandwidth compression is achieved without a proportional increase in the temporal record length. 

The proposed anamorphic transformation can be employed to engineer the modulation bandwidth of an ultrafast signal to match the sampling rate of the ADC while minimizing the number of samples needed to represent it. This physics-based grooming of the analog signal allows a conventional ADC to perform variable resolution sampling. The net result is that more samples are allocated to higher frequencies, where they are needed, and less to lower frequencies, where they are redundant.\\

\section*{Acknowledgements}

We would like to acknowledge valuable discussions with Dr. George Valley, Ata Mahjoubfar, Brandon Buckley and Peter DeVore. M. H. Asghari was supported by a Canadian NSERC. The work was partially supported by the NSF CIAN Engineering Research Center. \\

\appendix
\section{Table of parameters and acronyms}

\begin{tabular}{  >{\centering\arraybackslash}p{2.5cm} | >{\centering\arraybackslash}p{5cm}  }
 \multicolumn{2}{c}{} \\  
\multicolumn{2}{c}{Table 1. Parameters and acronyms used in this paper} \\
  \hline               
$t$	&Time  \\
\hline  
$\omega$	&Carrier Frequency\\
\hline
$\omega_m$	&Modulation intensity frequency, i.e. sideband frequency measured with respect to carrier frequency, $\omega$ \\
\hline
$\omega_s$ &	Digitizer sampling rate\\
\hline
$\Delta\omega_m$	&Modulation intensity bandwidth\\
\hline
$M$	&Modulation intensity bandwidth compression factor \\
\hline	
$N$	& Number of samples (Discrete-time record length)	\\
\hline
$E$ &Complex amplitude in time domain\\
\hline 
$\tilde{E}$&Electric field spectrum\\
\hline
$H(\omega)$	&Filter transfer function	\\
\hline
$\beta(\omega)$	&Filter phase response\\
\hline
GVD	&Group Velocity Dispersion\\
\hline
$\beta_2$	&Total 2nd order dispersion (GVD) coefficient\\
\hline
$\tau(\omega)$	&Group delay profile\\
\hline
$I$	&Intensity\\
\hline
FT	&Fourier Transform	\\
\hline
ADC	&Analog to digital converter	\\
\hline
AST&	Anamorphic Stretch Transform\\
\hline
MID	&Modulation Intensity Distribution	\\
\hline
DFT&	Time-stretch Dispersive Fourier Transform	\\
\hline

\end{tabular}


\begin{thebibliography}{99}


\bibitem{osastyle}	Y. Han, and B. Jalali, "Photonic time-stretched analog-to-digital converter: Fundamental concepts and practical considerations," Journal of Lightwave Technology 21, 3085-3103 (2003).
\bibitem{osastyle}  A. M. Fard, S. Gupta, and B. Jalali, "Photonic time-stretch digitizer and its extension to real-time spectroscopy and imaging," Laser and Photonics Reviews 7, 207-263 (2013).
\bibitem{osastyle}	K. Goda, and B. Jalali, "Dispersive Fourier transformation for fast continuous single-shot measurements," Nature Photonics 7, 102-112 (2013).
\bibitem{osastyle}	G. C. Valley, "Photonic analog-to-digital converters," Optics Express 15, 1955-1982 (2007).
\bibitem{osastyle}	A. Khilo, S. J. Spector, M. E. Grein, A. H. Nejadmalayeri, C. W. Holzwarth, M. Y. Sander, M. S. Dahlem, M. Y. Peng, M. W. Geis, N. A. DiLello, J. U. Yoon, A. Motamedi, J. S. Orcutt, J. P. Wang, C. M. Sorace-Agaskar, M. A. Popovic, J. Sun, G. Zhou, H. Byun, J. Chen, J. L. Hoyt, H. I. Smith, R. J. Ram, M. Perrott, T. M. Lyszczarz, E. P. Ippen, and F. X. Kartner, "Photonic ADC: Overcoming the bottleneck of electronic jitter," Optics Express 20, 4454-4469 (2012).
\bibitem{osastyle}	J. Stigwall and S. Galt, "Signal reconstruction by phase retrieval and optical backpropagation in phase-diverse photonic time-stretch systems," Journal of Lightwave Technology 25, 3017-3027 (2007). 
\bibitem{osastyle}	W. Ng, T. D. Rockwood, G. A. Sefler, and G. C. Valley, "Demonstration of a large stretch-ratio (M=41) photonic analog-to-digital converter with 8 ENOB for an input signal bandwidth of 10 GHz," IEEE Photonics Technology Letters 24, 1185-1187 (2012).
\bibitem{osastyle}	E. J. Candes and M. B. Wakin, "An introduction to compressive sampling," IEEE Signal Processing Magazine 25, 21–30 (2008).
\bibitem{osastyle} G. C. Valley, G. A. Sefler, T. J. Shaw, "Compressive sensing of sparse radio frequency signals using optical mixing," Optics Letters 37, 4675-4677 (2012).
\bibitem{osastyle} E. D. Diebold, N. K. Hon, Z. Tan, J. Chou, T. Sienicki, C. Wang and B. Jalali, “Giant tunable optical dispersion using chromo-modal excitation of a multimode waveguide,” Optics Express 19, 23809-23817 (2011).
\bibitem{osastyle} J. L. Hunt, B. G. Nickel, and C. Gigault, " Anamorphic images," American Journal of Physics 68, 232-237 (2000).
\bibitem{osastyle}	M. H. Asghari and B. Jalali, “Anamorphic transformation and its application to time-bandwidth compression,” Applied Optics, in Press.
\bibitem{osastyle}	M. H. Asghari and B. Jalali, "Warped dispersive transform and its application to analog bandwidth compression," IEEE Photonic Conference (IPC 2013), paper TUG 1.1, Sep 2013.
\bibitem{osastyle}	M. H. Asghari and B. Jalali, "Anamorphic Time Stretch Transform and its Application to Analog Bandwidth Compression," Accepted for 2013 IEEE GlobalSIP Symposium.
\bibitem{osastyle}	P. M. Woodward, Probability and information theory, with applications to Radar, Pergamon Press, New York, 1953.
\bibitem{osastyle}	M. H. Asghari, and B. Jalali, "Stereopsis-inspired time-stretched amplified real-time spectrometer (STARS)," IEEE Photonics Journal 4, 1693-1701 (2012).
\bibitem{osastyle} D. R. Solli, S. Gupta, and B. Jalali, “Optical phase recovery in the dispersive Fourier transform,” Applied Physics Letters 95, 231108 (2009).
\bibitem{osastyle} M. H. Asghari and J. Azana, "Self-referenced temporal phase reconstruction from intensity measurements using causality arguments in linear optical filters", Optics Letters 37, 3582-3584 (2012).
\bibitem{osastyle} F. Li, Y. Park, and J. Azana, "Linear characterization of optical pulses with durations ranging from the picosecond to the nanosecond regime using ultrafast photonic differentiation," Journal of Lightwave Technology 27, 4623–4633 (2009).
\bibitem{osastyle} C. Dorrer and I. Kang, "Complete temporal characterization of short optical pulses by simplified chronocyclic tomography," Optics Letters 28, 1481-1483 (2003).
\bibitem{osastyle} K. Goda, D. R. Solli, K. K. Tsia, and B. Jalali, “Theory of amplified dispersive Fourier transformation,” Physical Review A 80, 043821 (2009).
\bibitem{osastyle} D. R. Solli, J. Chou, and B. Jalali, "Amplified wavelength-time transformation for real-time spectroscopy," Nature Photonics 2, 48-51 (2008).
\bibitem{osastyle} D. R. Solli, C. Ropers, P. Koonath, and B. Jalali, "Optical rogue waves," Nature 450, 1054-1057 (2007).
\bibitem{osastyle} B. Wetzel, A. Stefani, L. Larger, P. A. Lacourt, J. M. Merolla, T. Sylvestre, A. Kudlinski, A. Mussot, G. Genty, F. Dias and J. M. Dudley, "Real-time full bandwidth measurement of spectral noise in supercontinuum generation," Scientific Reports 2, Article number: 882 (2012).
\bibitem{osastyle} K. Goda, K. K. Tsia, and B. Jalali, "Serial time-encoded amplified imaging for real-time observation of fast dynamic phenomena," Nature 458, 1145-1149 (2009).
\bibitem{osastyle} F. Qian, Q. Song, E. Tien, S. K. Kalyoncu, O. Boyraz, "Real-time optical imaging and tracking of micron-sized particles," Optics Communications 282, 4672–4675 (2009). 
\bibitem{osastyle} C. Zhang, Y. Qiu, R. Zhu, K. K. Y. Wong, and K. K. Tsia, "Serial time-encoded amplified microscopy (STEAM) based on a stabilized picosecond supercontinuum source," Optics Express 19, 15810-15816 (2011).
\bibitem{osastyle} M. A. Foster, R. Salem, D. F. Geraghty, A. C. Turner-Foster, M. Lipson and A. L. Gaeta, "Silicon-chip-based ultrafast optical oscilloscope," Nature 456, 81-84 (2008).
\bibitem{osastyle} T. Erdogan,  "Fiber grating spectra", Journal of Lightwave Technology 15,  1277-1294 (1997).
\bibitem{osastyle} M. G. F. Wilson and M. C. Bone, "Theory of curved diffraction gratings", Workshop on Integrated Optics, Technical University Berlin, 85-111, May 1980.
\bibitem{osastyle} G. A. Sefler and G. C. Valley, "Mitigation of group-delay-ripple distortions for use of chirped fiber-Bragg gratings in photonic time-stretch ADCs," Journal of Lightwave Technology 31, 1093-1100 (2013).


\end{thebibliography}
\end{document}